\documentclass[12pt,draftcls,onecolumn]{IEEEtran}
\usepackage[dvipdfmx]{graphicx,color}

\usepackage{amsmath,amsfonts,amssymb,amsthm}
\usepackage{latexsym}
\usepackage{bm}
\usepackage{color}

\newcommand{\cm}[1]{{\cal #1}}
\DeclareMathOperator{\diag}{diag}
\DeclareMathOperator{\trace}{trace}

\newtheorem{proposition}{Proposition}

\newtheorem{lemma}{Lemma}

\title{Min-Max Design of Feedback Quantizers for Netorwked Control Systems}

\author{Shuichi Ohno$^1$, Yuma Ishihara$^2$, and Masaaki Nagahara$^3$\\
$^{1}$~Hiroshima University, 
1-4-1 Kagamiyama, Higashi-Hiroshima, 739-8527, JAPAN\\
$^{2}$~JTEKT Corporation \\ 
$^{3}$~
Institute of Environmental Science and Technology,
The University of Kitakyushu,
Hibikino 1-1, Wakamatsu-ku, Kitakyushu, Fukuoka
808-0135, JAPAN
}

\begin{document}

\maketitle

\begin{abstract}
In a networked control system, quantization is inevitable to transmit 
control and measurement signals. 
While uniform quantizers are often used in practical systems,
the overloading, which is 
due to the limitation on the number of bits in the quantizer,
may significantly degrade the control performance. 
In this paper, we design an overloading-free feedback quantizer based on 
a $\Delta\Sigma$ modulator,
composed of an error feedback filter and a static quantizer.
To guarantee no-overloading in the quantizer, 
we impose an $l_{\infty}$ norm constraint on the feedback signal 
in the quantizer.  
Then, for a prescribed $l_{\infty}$ norm constraint on the error 
at the system output induced by the quantizer, 
we design the error feedback filter that requires 
the minimum number of bits that achieves the constraint. 
Next, for a fixed number of bits for the quantizer, 
we investigate the achievable minimum $l_{\infty}$ norm of 
the error at the system output with an overloading-free quantizer. 
Numerical examples are provided to validate our analysis and synthesis. 
\end{abstract}

\begin{IEEEkeywords}
   quantization, overloading, delta-sigma modulation,
networked control, linear matrix inequalities
\end{IEEEkeywords}

\section{Introduction}
\label{sec:introduction}
In a networked (or distributed) control system,
multiple geographically distributed systems
exchange their information to achieve control tasks.
For example, sensors at a controlled plant
send their observation signals
to a controller, and the controller transmits control signals 
to actuators at the plant through communication channels
(e.g., see \cite{1299533} and the reference therein). 

If distributed systems are connected by reliable communication channels, 
then a sufficient level of accuracy of data transmission can be assured. 
However, it is often the case that communication rates 
are limited due to physical constraints 
especially when wireless communication is used. 
To transmit signals over rate-limited digital communication channels,  
the continuous-valued (or even discrete-valued) signals
have to be quantized into low-resolution signals.
When only a small number of bits can be assigned to represent the signals, 
quantization errors may cause serious degradation in the stability and control performance. This motivates the research on control under limited data rates. 

The minimum data rate to keep the state of 
a closed-loop system in a bounded region with state feedback control
has been provided in \cite{wong1999systems}. 
Stabilizability and observability under a communication 
constraint has been studied in \cite{Tatikonda2004} 
for discrete-time, linear and time-invariant (LTI) systems. 
In \cite{Tatikonda2004}, a sufficient and necessary condition on the information rate for 
asymptotic stabilizability and observability has also been presented. 
The same condition has been shown in \cite{Nair2003} for
a sufficient and necessary condition for 
exponentially stabilizability of 
discrete-time LTI systems with random initial values. 
The discussions on the minimum data rate based on the information theory
gives valuable insights into control under limited data rates. 
However, even if the rate is assured, 
the closed-loop system cannot be always stabilized in practice,
since the minimum data rate is not a constant rate 
for each time slot but an averaged  rate over time. 
Also, the rate is evaluated under the ideal assumption that 
the channel from the controller to the plant has infinite precision 
and the quantizer has an infinite range for its input. 
Moreover, the minimum rate is attained by a very complicated
quantizer that is hard to implement.
In practice, the control signal should be bounded 
in a fixed range due to physical requirements,
only a finite number of bits can be assigned, 
and  the quantizer has a limited range. 
Taking account of these limitations, 
we develop an implementable quantizer  
that requires only a small number of bits 
for quantization to attain requirements on control performance. 

Quantization with error feedback has been originally 
developed to reduce quantization error in the coefficients of 
digital filters \cite{1162925,1163989,134473}, 
where the quantization error of a static uniform quantizer 
is filtered by an  error feedback filter 
and then it is fed back to the input to the static uniform quantizer 
(see Fig. \ref{fig:efb_quantizer} in Section II).  
On the other hand, $\Delta \Sigma$ modulators also employ the
error feedback mechanism, and are often utilized in practice to convert 
real numbers into fixed-point numbers \cite{Delta-Sigma}, 
since they can be implemented at a relatively low cost.  

For networked control systems, 
a variant of $\Delta \Sigma$ modulator has been studied 
in \cite{Azuma2008b}, which is called a {\it dynamic quantizer}.
The parameters in the dynamic quantizer can be obtained by
linear programming \cite{Azuma2008} 
and by convex optimization \cite{Sawada2010c}. 
To avoid overloading, \cite{Azuma2008} proposes to limit the $l_{\infty}$ norm 
of the feedback signals.
However, the dynamic quantizer only supports 
a smaller set of error feedback filters
than conventional $\Delta\Sigma$ modulators
and hence the optimal performance cannot be guaranteed \cite{7769239}. 

Recently, $H_\infty$ optimal design of error feedback filters has been proposed based on the generalized Kalman-Yakubovich-Popov lemma \cite{6156470, li2014design}. 
Also, a post filter connected to the $\Delta\Sigma$ modulator is 
incorporated into the design of the error feedback filter \cite{6461108}   
and the weighted noise spectrum is also exploited \cite{6617704}.
In \cite{4522532}, under the assumption that the output of 
the static uniform quantizer is a white noise, 
the optimal error feedback filter has been synthesized 
such that it minimizes the variance of the quantization error subject to 
the constraint on the variance of the input to the static uniform quantizer. 
However, the constraint on the variance 
does not necessarily guarantee no-overloading in the quantizer.  
In practical control systems such as nuclear plants,
an overloading may cause instability followed by a catastrophic accident.  
To assure that no overloading occurs, 
we should take into account the maximum absolute value, i.e.,  
the $l_{\infty}$ norm of the input to the  quantizer.

This paper develops a quantizer with error feedback 
that needs a small number of bits required for quantization to achieve 
the requirement on the worst-case error in the control output, 
while keeping no-overloading in the quantizer. 
We regulate the $l_{\infty}$ norm of the feedback signals in the quantizer 
to assure no-overloading.  
In our preliminary conference paper \cite{ohno2017eusipco}, 
the IIR feedback filter is adopted to minimize 
the $l_{\infty}$ norm of the error at the output 
under the constraint on the $l_{\infty}$ norm of the feedback signals. 
In the study of \cite{ohno2017eusipco}, 
an upper bound of the $l_{\infty}$ norm, which is not tight, was utilized, 
since the exact $l_{\infty}$ norm of an IIR filter is not easily evaluated. 
Alternatively, we here adopt FIR feedback filters since
the $l_{\infty}$ norm of an FIR filter can be exactly and directly evaluated.
We formulate the design of the optimal FIR feedback filter 
as linear programming, which can be readily solved numerically.  
The minimum number of bits assigned to the quantizer is 
determined with the optimized feedback filters. 
As illustrated by our numerical results,  
the optimized FIR feedback filter show a better performance 
than the IIR feedback filter proposed in \cite{ohno2017eusipco}.  
Next,  for a given number of bits for the quantizer, 
we investigate the achievable minimum $l_{\infty}$ norm of 
the error at the system output induced by the quantizer 
with an overloading-free quantizer. 
This can be enabled by finding the relationship between  
$l_{\infty}$ norm of the feedback signal 
and the $l_{\infty}$ norm of the error in the output, 
which can be obtained by solving convex optimization problems. 
Numerical examples are provided to validate our analysis and synthesis. 

 This paper is organized as follows. 
 Networked systems and quantization are reviewed in Section II. 
Then, quantizers are synthesized in Section III
based on the $l_{\infty}$ norm of the effect of the quantization error 
and the output of the error feedback filter. 
Section IV presents numerical results on our synthesis 
 and Section V concludes this paper. 

\subsection*{Notation}
$\mathbb{Z}$,
$\mathbb{R}$, and
$\mathbb{R}_{+}$
stand for the set of real numbers,
integers, and non-negative real numbers,
respectively.
The $\mathrm z$ transform of a sequence $f=\{f_k\}_{k=0}^{\infty}$ 
is denoted as $F[\mathrm z]=\sum_{k=0}^{\infty}f_k \mathrm z^{-k}$.
The output sequence $h$ of an linear and time-invariant (LTI)
system $F[\mathrm z]$ with the input sequence $g$
(i.e. $h=f\ast g$ where $\ast$ denotes the convolution)
is expressed as $h=F[\mathrm z]g$.
The $l_{\infty}$ norm of a scalar-valued sequence $x=\{x_k \}$ is defined as 
$||x||_{\infty}=\max_{k} |x_k|$. 

\section{Error Feedback Quantizer}

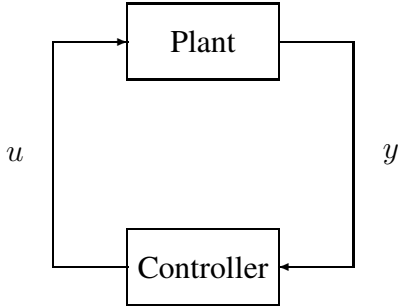
\begin{figure}[tbp]\setlength{\unitlength}{10mm}
  \centering
 \begin{center}
\leavevmode
\begin{picture}(6,5)(0,0)
\put(2,0.5){\framebox(2,1){Controller}}
\put(2,3.5){\framebox(2,1){Plant}}
\put(4,4){\line(1,0){1}}
\put(5,4){\line(0,-1){3}}
\put(5,1){\vector(-1,0){1}}
\put(2,1){\line(-1,0){1}}
\put(1,1){\line(0,1){3}}
\put(1,4){\vector(1,0){1}}
\put(0.5,2.5){\makebox(0,0)[c]{$u$}}
\put(5.5,2.5){\makebox(0,0)[c]{$y$}}
\end{picture}
\end{center}
\caption{Feedback control system.}
\label{fig:plant}
\end{figure}

Fig. \ref{fig:plant} depicts a feedback control system,
in which the plant is assumed to be  linear and time-invariant (LTI), and 
the signals $y$ and $u$ are 
functions of time in general. 
Based on the observation signal $y$ from the plant, 
the controller generates the control input $u$ to the plant.  

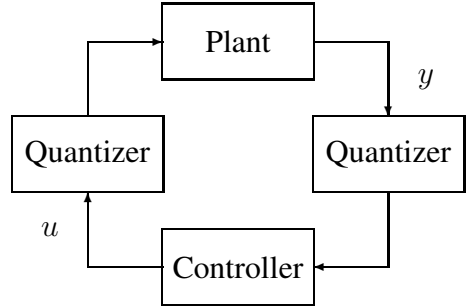
\begin{figure}[tbp]\setlength{\unitlength}{10mm}
  \centering
 \begin{center}
\leavevmode
\begin{picture}(6,5)(0,0)
\put(2,0.5){\framebox(2,1){Controller}}
\put(2,3.5){\framebox(2,1){Plant}}
\put(4,4){\line(1,0){1}}
\put(5,4){\vector(0,-1){1}}
\put(4,2){\framebox(2,1){Quantizer}}
\put(5,2){\line(0,-1){1}}
\put(5,1){\vector(-1,0){1}}
\put(2,1){\line(-1,0){1}}
\put(1,1){\vector(0,1){1}}
\put(0,2){\framebox(2,1){Quantizer}}
\put(1,3){\line(0,1){1}}
\put(1,4){\vector(1,0){1}}
\put(0.5,1.5){\makebox(0,0)[c]{$u$}}
\put(5.5,3.5){\makebox(0,0)[c]{$y$}}
\end{picture}
\end{center}
\caption{Control system with quantization.}
\label{fig:system_ruantizer}
\end{figure}

The observation signal $y$ and 
the control input $u$ are assumed to be 
transmitted through digital communication channels. 
If $y$ and $u$ are real-valued signals, 
quantization is required to convert them into discrete-valued signals 
before transmission as illustrated in Fig. \ref{fig:system_ruantizer}. 
Note that even if $y$ and $u$ are discrete-valued digital signals, 
they may have to be rounded off when 
the capacities of the communication channels are limited. 

\begin{figure}[tbp]\setlength{\unitlength}{10mm}
  \centering
 \begin{center}
\leavevmode
\begin{picture}(6,5)(0,0)
\put(2,0.5){\framebox(2,1){Controller}}
\put(2,3.5){\framebox(2,1){Plant}}
\put(4,4){\line(1,0){1}}
\put(5,4){\vector(0,-1){1.4}}
\put(5,2.5){\circle{0.2}}
\put(5,2.5){\makebox(0,0){\small +}}
\put(6,2.5){\vector(-1,0){0.9}}
\put(5,2.4){\line(0,-1){1.4}}
\put(5,1){\vector(-1,0){1}}
\put(2,1){\line(-1,0){1}}
\put(1,1){\vector(0,1){1.4}}
\put(1,2.5){\circle{0.2}}
\put(1,2.5){\makebox(0,0){\small +}}
\put(0,2.5){\vector(1,0){0.9}}
\put(1,2.6){\line(0,1){1.4}}
\put(1,4){\vector(1,0){1}}
\put(0.5,1.5){\makebox(0,0)[c]{$u$}}
\put(5.5,3.5){\makebox(0,0)[c]{$y$}}
\put(-0.2,2.5){\makebox(0,0)[c]{$e_c$}}
\put(6.3,2.5){\makebox(0,0)[c]{$e$}}
\end{picture}
\end{center}
\caption{Control system and quantization error signals.}
\label{fig:system_error}
\end{figure}
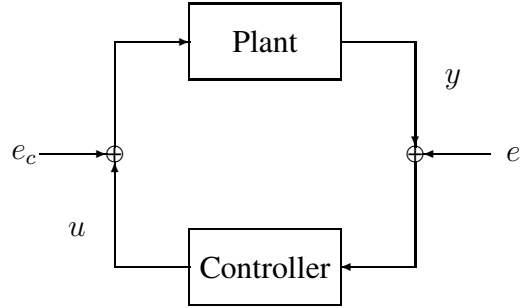

The difference between the input and the output of the quantizer is called
the quantization error.
There are two quantization errors;
one is the quantization error denoted by $e_c$ for  the control signal $u$
and the other is
the quantization error $e$ for the observation signal $y$. 
With these quantization errors, 
the control system in Fig. \ref{fig:system_ruantizer} can be modeled
by an additive-noise control system shown in Fig. \ref{fig:system_error}. 

Controllers are often connected to plants through wired networks. 
On the other hand, the observation signal are collected by sensors,  
which may be connected through wireless networks. 
Thus, we here focus on the quantization error $e$ of 
the observation signal, 
assuming that there is no quantization error at the controller.  
We also assume that each sensor observes a scalar-valued signal to be quantized 
and works independently of the other sensors. 
Since we consider the independent quantization of each entries of $y$, 
we assume $y$ to be a scalar-valued signal to a particular quantizer  
for simplicity of presentation.   
We also assume the plant is a
single-input and single-output (SISO) system.  
Note that, most of our results may be applied 
to the quantization error at the controller 
and the multiple-input and multiple-output (MIIMO) systems.  
We assume the reachability and the observability of the plant,
without which the plant cannot be stabilized in general.

In quantization, real numbers are mapped into their binary representation. 
Fixed-point representation and floating-point representation are available
for quantization.  
In this paper, 
we take fixed-point representation into account since it is often adopted 
in embedded systems. 

Let us take a static uniform quantizer for example.  
The static uniform quantizer can be described by two parameters, 
the quantization interval $d \in \mathbb{R}_{+}$ 
and the saturation level $L \in \mathbb{R}_{+}$. 
For simplicity, we assume that $L$ is an integer multiple  of $d$. 
For the static quantizer, let us consider a mid-rise quantizer \footnote{Similar results can be obtained for mid-tread quantizers with slight modifications.} 
 $Q(\xi)$ expressed as 
\begin{equation}
  \label{eq:1}
  Q(\xi)=
\left\{
  \begin{array}{ll}
    \left(
      i+\frac{1}{2}
    \right)     d, &|\xi|\leq L+\frac{d}{2} \\
    & \quad \quad
    \mbox{~~and~~} \xi \in [id, (i+1)d), i \in {\mathbb{Z}}\\
    L,& \xi >  L +\frac{d}{2}\\
    -L, &\xi <  -L-\frac{d}{2} \\
  \end{array}
\right.
\end{equation}
The overloading is the saturation due to the fixed number of bits 
to represent the quantized values in binary. 
For the mid-rise quantizer, 
the overloading occurs if $|\xi| > L+\frac{d}{2}$. 

The static uniform quantizer is often utilized in practice but   
its errors and effects of the overloading are not negligible  
unless a sufficient number of bits is assigned to the quantizer. 
To mitigate these influences,
we adopt a quantizer with an error feedback filter,  
which has been originally developed to reduce the effects 
of the quantized coefficients in digital filters \cite{1162925,1163989,134473}. 

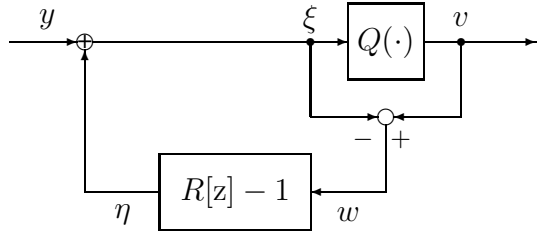
\begin{figure}[tbp]\setlength{\unitlength}{10mm}
  \centering
 \begin{center}
\leavevmode
\begin{picture}(7,3.5)(0,-0.5)
\put(0,2){\vector(1,0){0.9}}
\put(1,2){\circle{0.2}}
\put(1,2){\makebox(0,0){\small +}}
\put(1.1,2){\vector(1,0){3.4}}
\put(4,2){\circle*{0.1}}
\put(4.5,1.5){\framebox(1,1){$Q(\cdot)$}}
\put(5.5,2){\vector(1,0){1.5}}
\put(6,2){\circle*{0.1}}
\put(4,2){\line(0,-1){1}}
\put(6,2){\line(0,-1){1}}
\put(4,1){\vector(1,0){0.9}}
\put(6,1){\vector(-1,0){0.9}}
\put(5,1){\circle{0.2}}
\put(5,0.9){\line(0,-1){0.9}}
\put(5,0){\vector(-1,0){1}}
\put(2,-0.5){\framebox(2,1){$R[\mathrm z]-1$}}
\put(2,0){\line(-1,0){1}}
\put(1,0){\vector(0,1){1.9}}
\put(4.5,-0.3){\makebox(0,0)[c]{$w$}}
\put(1.5,-0.3){\makebox(0,0)[c]{$\eta$}}
\put(0.5,2.3){\makebox(0,0)[c]{$y$}}
\put(4,2.3){\makebox(0,0)[c]{$\xi$}}
\put(6,2.3){\makebox(0,0)[c]{$v$}}
\put(5.2,0.7){\makebox(0,0)[c]{\small $+$}}
\put(4.7,0.7){\makebox(0,0)[c]{\small $-$}}
\end{picture}
\end{center}
\caption{Error feedback quantizer.}
\label{fig:efb_quantizer}
\end{figure}

Fig. \ref{fig:efb_quantizer} illustrates a block diagram 
of our quantizer. 
The quantization error, or the round-off error, 
of the static uniform quantizer $Q(\cdot)$ is defined as 
\begin{equation}
  \label{eq:2}
    w_{k}=v_{k}-\xi_{k},\quad k=0,1,2,\ldots
\end{equation}
where $\xi_{k}$ and $v_{k}$ are respectively the input and the output vectors 
of the static uniform quantizer at time $k$.
Note that the round-off error $w$ of the static quantizer 
is different from the quantization error defined as 
\begin{equation}
  \label{eq:50}
e=v-y.   
\end{equation}

The round-off error signal $w$ is filtered by the error feedback filter 
$R[\mathrm z]-1$ and then it is 
fed  back to the input to the quantizer. 
The error feedback filter $R[\mathrm z]-1$ has to be strictly proper, 
that is, $R[\infty]=1$.   
The quantizer in Fig.~\ref{fig:efb_quantizer} is also known as 
a $\Delta\Sigma$ modulator, which is an efficient analog to digital
 (A/D) converter with feedback from the output 
of a static uniform quantizer to shape the quantization noise \cite{Delta-Sigma}. 
In the $\Delta\Sigma$ modulator, $R[\mathrm z]$  
is called a noise shaping filter or a noise transfer function. 

The input signal $\xi$ to the static quantizer can be expressed 
from Fig. \ref{fig:efb_quantizer} as 
\begin{equation}
  \label{eq:3}
  \xi=y+(R[\mathrm z]-1)w.
\end{equation}
From \eqref{eq:2} and \eqref{eq:3},
the quantization error $e$ is given by 
\begin{equation}
  \label{eq:5}
  e=v-y=R[\mathrm z]w.
\end{equation}
Thus, the output signal of the quantizer can be expressed as 
\begin{equation}
  \label{eq:6}
  v=y+e=y+R[\mathrm z]w.  
\end{equation}

Let the  output signal of interest be $z$  
and the transfer function from $y$ to $z$ 
be $H[\mathrm z]$.  
Then, since the plant is assumed to be reachable and observable, there exists 
a controller that stabilizes the control system 
when there is no quantization error. 
With this controller, $H[\mathrm z]$ is stable.  
Thus, without loss of generality, we can assume 
that $H[\mathrm z]$ is stable.  

Since $e$ also goes through $H[\mathrm z]$, 
the error signal in $z$ that comes from 
the quantization error $e$ can be expressed as 
\begin{equation}
  \label{eq:8}
  \epsilon=H[\mathrm z]e=H[\mathrm z] R[\mathrm z]w
.
\end{equation}
Unless $H[\mathrm z] R[\mathrm z]=0$,  
we cannot assure that $\epsilon_k \rightarrow 0$ as 
$k \rightarrow \infty$ due to the unpredictable round-off errors.
Thus, we cannot guarantee the exponential stability of the feedback system,  
even if $H[\mathrm z]$ is stable.   
All we can do is to mitigate the effect of the quantization 
given by \eqref{eq:8}. 
Thus, our goal is to design 
an error feedback quantizer so that
the maximum absolute value of $\epsilon$ is not greater 
than a prescribed threshold $\gamma_{\epsilon}$, 
which can be expressed as 
\begin{equation}
  \label{eq:510}
  \max_{k} |\epsilon_k| \leq \gamma_{\epsilon} 
\end{equation}
or equivalently as 
\begin{equation}
\label{eq:20}
  ||\epsilon||_{\infty} \leq \gamma_{\epsilon}.  
\end{equation}

\section{Synthesis of Error Feedback Quantizer }
\label{sec:trad-betw-round}
Unless overloading occurs, the round-off error $w$ is bounded 
such as 
\begin{equation}
  \label{eq:530}
  ||w||_{\infty} \leq \frac{d}{2}. 
\end{equation}
Otherwise, the signal $z$ of interest 
cannot be bounded in general, 
since $w$ may be unbounded due to overloading. 
Then, the overloading complicates the control law, 
since $||\epsilon||_{\infty}$ depends on it. 
To design the controller and the quantizer independently,   
it is better to avoid the overloading in the static uniform quantizer. 

If there is no overloading, then $y$ is bounded, 
since the system is stable and the error $\epsilon$ is bounded 
with a stable $R[\mathrm z]$. 
Without loss of generality, we can assume that 
the observation signal has the symmetric magnitude limitation described as
\begin{equation}
  \label{eq:13}
  ||y||_{\infty} \leq L_{y}
  . 
\end{equation}

Let us adopt the static uniform quantizer characterized 
by \eqref{eq:1} in our error feedback quantizer. 
From our definitions, if the feedback signal $\eta$ meets 
\begin{equation}
  \label{eq:15}
||y+ \eta||_{\infty} \leq L+\frac{d}{2}
,
\end{equation}
then any overloading never happens at the static quantizer. 

Let us introduce the norm of a system 
$F[\mathrm z]=\sum_{k=0}^{\infty}f_k{\mathrm z}^{-k}$ 
induced by the $l_\infty$ norm of the input and output signals
defined as \cite{832826} 
\begin{equation}
  \label{eq:16}
  ||F[\mathrm z]||=\sup_{x \ne 0}\frac{||F[\mathrm z]x||_\infty}{||x||_\infty},  
\end{equation}
for $x\in l_\infty$ (i.e. $\|x\|_\infty<\infty$).
If $F[\mathrm z]$ is an SISO system, 
the norm is equivalent to the $l_1$ norm of the impulse response of the system, 
that is, we have 
\begin{equation}
  \label{eq:17}
  ||F[\mathrm z]||=\sum_{k=0}^{\infty} |f_k|. 
\end{equation}
It follows from \eqref{eq:15} and 
$||\eta||_{\infty}\leq ||R[\mathrm z]-1|| (d/2)$ that if one sets 
\begin{equation}
  \label{eq:18}
  L_{y} + ||R[\mathrm z]-1|| \frac{d}{2} \leq L+\frac{d}{2}
\end{equation}
then no-overloading in the static uniform quantizer is assured. 
In other words, the $l_{\infty}$ norm of the feedback signal 
should be equal to or less than $L+d/2 -L_{y}$. 

For the binary representation of the observation signals, 
we have to determine its accuracy and range, i.e., 
the quantization interval $d$ 
and the saturation level $L$ for the uniform quantizer. 
If we assign $b$ bits to represent the observation signal,  
we have  
\begin{equation}
  \label{eq:14}
  2 L= (2^b-1) d.
\end{equation}
From \eqref{eq:18} and \eqref{eq:14}, we summarize the above discussion as a proposition:
\begin{proposition}
There is no overloading in the error feedback quantizer 
composed of an error feedback filter 
$R[\mathrm z]-1$ and a mid-rise quantizer if 
\begin{equation}
  \label{eq:19}
  L_y+||R[\mathrm z]-1|| \frac{d}{2} \leq  2^{b-1} d
\end{equation}
where $b$ is the number of bits assigned to the mid-rise quantizer, 
$d$ and $L_y$  denote its  quantization interval and 
saturation level respectively.    
\end{proposition}

Now, we would like to find the number of bits that assures 
no-overloading under the constraint \eqref{eq:20}, 
which is, from \eqref{eq:8}, achieved if 
\begin{equation}
  \label{eq:21}
  ||H[\mathrm z]R[\mathrm z]||\frac{d}{2} \leq  \gamma_{\epsilon}
.
\end{equation}
To obtain the minimum $b$ that satisfies \eqref{eq:19}, 
we set the quantization interval of our static uniform quantizer 
to be 
\begin{equation}
  \label{eq:22}
  d=\frac{2 \gamma_{\epsilon}}{||H[\mathrm z]R[\mathrm z]||}
.
\end{equation}
Substituting \eqref{eq:22} into \eqref{eq:19} leads to 
\begin{equation}
  \label{eq:23}
  \frac{L_y}{\gamma_{\epsilon}}
||H[\mathrm z]R[\mathrm z]||+||R[\mathrm z]-1|| \leq 
2^{b}. 
\end{equation}

For given $L_y$ and $\gamma_{\epsilon}$, 
the left hand side of the inequality above 
can be evaluated with $||H[\mathrm z]R[\mathrm z]||$ and 
$||R[\mathrm z]-1||$, whose minimum can be obtained by 
solving the following optimization problem:
\begin{equation}
  \label{eq:24}
  \min_{R[\mathrm z] \in RH_{\infty},\gamma_{\epsilon}}  
  c \tilde{\gamma}_{\epsilon}+\tilde{\gamma}_{\eta}
\end{equation}
subject to $R[\infty]=1$ and 
\begin{align}
  \label{eq:25}
  ||H[\mathrm z]R[\mathrm z]|| & \leq \tilde{\gamma}_{\epsilon}\\
  \label{eq:26}
  ||R[\mathrm z]-1|| & \leq \tilde{\gamma}_{\eta}
\end{align}
where $RH_{\infty}$ is the set of 
stable proper rational functions with real coefficients 
and 
\begin{equation}
  \label{eq:540}
 c=\frac{L_y}{\gamma_{\epsilon}}.  
\end{equation}
It should be noted that 
the objective function is a linear in $\tilde{\gamma}_{\epsilon}$ 
and $\tilde{\gamma}_{\eta}$. 

The problem above can be solved if we restrict $R[\mathrm z]$ to 
have a finite impulse response (FIR). 
On the other hand, the global optimal solution 
is not available for general infinite impulse response (IIR) filters.  
 
\subsection{FIR filter design}
\label{section:FIR}
If $R[\mathrm z]$ is an FIR filter of order $n$, 
then the problem can be formulated as a linear programming (LP) 
and be numerically solved as follows. 

To solve the problem, 
we express the composite system $H[\mathrm z]R[\mathrm z]$ 
as a state-space realization. 
We denote the state-space matrices 
of a state-space realization of $H[\mathrm z]$ as $(A_h, B_h, C_h, D_h)$, 
while the state-space matrices  
of a state-space realization of $R[\mathrm z]$ as $(A_r, B_r, C_r, D_r)$ 
with $D_r=1$.  
Then, the state-space realization of $H[\mathrm z]R[\mathrm z]$  
can be written as 
\begin{eqnarray}
  \label{eq:27}
  x_{k+1}&=&\cm{A}x_{k}+\cm{B} w_{k}\\
  \epsilon_{k}&=& \cm{C}x_{k}+\cm{D}w_{k}
\end{eqnarray}
where the state-space matrices for this are given as
\begin{eqnarray}
  \label{eq:28}
  \cm{A}&=&
  \left[
    \begin{array}{cc}
      A_{r} & B_{r} C_h\\
      \bm{0} & A_h 
    \end{array}
  \right]
\\
  \cm{B}&=&
  \left[
    \begin{array}{c}
      B_{r}\\
       B_h 
    \end{array}
  \right]
\\
  \cm{C}&=&
  \left[
    \begin{array}{cc}
      C_{r} & D_{r} C_h
    \end{array}
  \right] \label{eq:cmC}
\\
  \cm{D}&=&D_h
.
\end{eqnarray}
The impulse response from $w$  to $\epsilon$ can be expressed as 
\begin{eqnarray}
  \label{eq:29}
  f_k=
  \left\{
    \begin{array}{cc}
      \cm{D}, & k=0\\
      \cm{C} \cm{A}^{k-1}\cm{B}, & k\ne 0
    \end{array}
  \right.
\end{eqnarray}

A state-space realization $(A_{r},B_{r},C_{r},1)$ of 
the FIR filter $R[\mathrm z]=1+\sum_{k=1}^{n}r_k \mathrm z^{-k}$ 
is given by 
\begin{align}
A_{r}&=
\begin{bmatrix}
0 & 1 & 0 & \cdots &  0\\
\vdots  & \ddots &\ddots & \ddots & \vdots\\
\vdots  &  & \ddots  & \ddots  & 0 \\
\vdots  &  &  & \ddots  & 1 \\
0 & \cdots & \cdots &\cdots & 0 \\  
\end{bmatrix}
, \ B_{r}=
\begin{bmatrix}
0 \\
\vdots \\
0 \\
1 \\
\end{bmatrix}
\\
C_{r}&=
\begin{bmatrix}
r_{n}, & r_{n-1}, & \cdots & r_{1} 
\end{bmatrix}, 
D_{r}=1.
\label{eq:CrDr}
\end{align}
Since $A_{r}$ and $B_{r}$ are constant, 
$\cm{A}$, $\cm{B}$, and $\cm{D}$ 
are constant.  
Moreover, $\cm{A}$ is Shur, that is, all eigenvalues of $\cm{A}$ are strictly
inside the unit circle, since $H[\mathrm z]$ is stable.   

For a sufficiently large integer $m$, 
we can approximate $||H[\mathrm z]R[\mathrm z]||$ such that 
\begin{equation}
  \label{eq:30}
  ||H[\mathrm z]R[\mathrm z]||=|D_h|+
\sum_{k=1}^{m} |\cm{C} \cm{A}^{k-1}\cm{B}|.
\end{equation}
Then, our problem can be expresses as the following minimization problem: 
\begin{equation}
  \label{eq:31}
  \min_{r_1,\ldots,r_n}    c \tilde{\gamma}_{\epsilon}+\tilde{\gamma}_{\eta}
\end{equation}
subject to 
\begin{align}
  \label{eq:32}
|D_h|+\sum_{k=1}^{m} |\cm{C} \cm{A}^{k-1}\cm{B}| & \leq \tilde{\gamma}_{\epsilon}\\
  \label{eq:33}
\sum_{k=1}^{n}|r_k| & \leq \tilde{\gamma}_{\eta}
.
\end{align}
Note that the matrix $\cm{C}$ depends linearly on $r_1,\ldots,r_n$ as in
\eqref{eq:cmC} and \eqref{eq:CrDr}.

Introducing non-negative auxiliary variables $\bar{f}_k$ 
for $k=1,\ldots,m$, 
one can express   
\eqref{eq:32} as
\begin{align}
  \label{eq:34}
&  \sum_{k=1}^{m} \bar{f}_k\leq  \tilde{\gamma}_{\epsilon}\\
  \label{eq:35}
& -\bar{f}_k \leq |D_h|+\cm{C} \cm{A}^{k-1}\cm{B} \leq \bar{f}_k 
\quad \mbox{for} \quad k=1,\ldots, m
\end{align}
Similarly, with non-negative auxiliary variables 
$\bar{r}_k \geq 0$ for $k=1,\ldots,n$, 
\eqref{eq:33} is equivalent to 
\begin{align}
  \label{eq:36}
&  \sum_{k=1}^{n} \bar{r}_k\leq  \tilde{\gamma}_{\eta}\\
  \label{eq:37}
& -\bar{r}_k \leq r_k \leq \bar{r}_k 
\quad \mbox{for} \quad k=1,\ldots, n
.
\end{align}
Then, our minimization problem is formulated as 
the following linear programming (LP):
\begin{equation}
  \label{eq:38}
  \min_{r_1,\ldots,r_n,\bar{f}_1,\ldots,\bar{f}_m,\bar{r}_1,\ldots,\bar{r}_n,}  
  c \tilde{\gamma}_{\epsilon}+\tilde{\gamma}_{\eta}
\end{equation}
subject to \eqref{eq:34}, \eqref{eq:35}, \eqref{eq:36}, 
\eqref{eq:37}, and 
\begin{align}
  \label{eq:39}
& \bar{f}_k \geq 0 
\quad \mbox{for} \quad k=1,\ldots, m\\
  \label{eq:40}
&\bar{r}_k \geq 0 
\quad \mbox{for} \quad k=1,\ldots, n
.
\end{align}

\subsection{IIR filter design}
\label{section:IIR}
Let us shortly introduce the IIR filter design proposed in 
\cite{ohno2017eusipco}, 
where the order of $R[\mathrm z]$ is set to be equal to the 
order of $H[\mathrm z]$. For the design of IIR filters, we re-express 
the state-space realization of $H[\mathrm z]R[\mathrm z]$ as 
\begin{eqnarray}
  \label{eq:128}
  \cm{A}&=&
  \left[
    \begin{array}{cc}
      A_h & B_h C_r\\
      \bm{0} & A_r 
    \end{array}
  \right]
\\
  \cm{B}&=&
  \left[
    \begin{array}{c}
      B_h\\
       B_r 
    \end{array}
  \right]
\\
  \cm{C}&=&
  \left[
    \begin{array}{cc}
      C_h & D_hC_r
    \end{array}
  \right]
.
\end{eqnarray}

In \cite{Shingin}, the following lemma has been provided 
by using the invariant set of a discrete-time system.
\begin{lemma}
Suppose the initial state $x_0$ is $0$ and 
the input $w$ is bounded as $||w||_{\infty} \leq 1$.
Then, the state vectors $x_k, k=1,2,\ldots$ remain in the ellipsoid
\begin{equation}
  \label{eq:129}
 \cm{E}(\cm{P})=\{x :x^T \cm{P} x \leq 1\}
\end{equation}
if and only if there exist a scalar $\alpha \in [0,1-\rho^2(\cm{A})]$ 
and a positive definite matrix $\cm{P}$ satisfying
\begin{equation}
  \label{eq:130}
        \left[
    \begin{array}{ccc}
      (1-\alpha)\cm{P} &\bm{0}&\cm{A}^T\cm{P}\\
      \bm{0} & \alpha  &\cm{B}^T\cm{P}\\
      \cm{P}\cm{A} &\cm{P}\cm{B}& \cm{P}
    \end{array}
  \right]
  \succeq \bm{0}
, 
\end{equation}
where $\rho(A)$ is the spectrum radius of $A$. 
\end{lemma}

It follows from $\epsilon_k=\cm{C}x_k+D_hw_k$ that 
if $x_k \in \cm{E}(\cm{P})$, then 
\begin{equation}
  \label{eq:131}
  \sup_{x_k \in \cm{E}(P)} |\epsilon_k-D_hw_k|^2 
\leq 
  \trace 
  \left(
\cm{C}\cm{P}^{-1}\cm{C}^T
  \right)
. 
\end{equation}
Thus, we have 
\begin{equation}
  \label{eq:132}
    ||\epsilon||_{\infty} \leq  
|D_h|\frac{d}{2}+
    \left[
  \trace 
  \left(
\cm{C}\cm{P}^{-1}\cm{C}^T
  \right)
    \right]^{\frac{1}{2}}
.
\end{equation}

On the other hand, with 
\begin{equation}
  \label{eq:133}
\tilde{\cm{C}} =
  \left[
    \begin{array}{cc}
      \bm{0} & C_r
    \end{array}
  \right]
\end{equation}
we can express $\eta$ as 
\begin{equation}
  \label{eq:134}
  \eta_k=\tilde{\cm{C}}x_k  
\end{equation}
which leads to 
\begin{equation}
  \label{eq:135}
    ||\eta||_{\infty} \leq 
    \left[
\trace
\left(
\tilde{\cm{C}}\cm{P}^{-1}\tilde{\cm{C}}^T
\right)
\right]^{\frac{1}{2}}
. 
\end{equation}
Unlike FIR filters, we cannot analytically 
evaluate $||\epsilon||_{\infty}$  and $||\eta||_{\infty}$.  
Instead, we  consider the minimization 
using the right hand sides of \eqref{eq:132} and \eqref{eq:135}, 
that is, the upper bounds of $||\epsilon||_{\infty}$  
and $||\eta||_{\infty}$, such that: 
\begin{align}
\label{eq:137}
  \trace 
  \left(
\cm{C}\cm{P}^{-1}\cm{C}^T
  \right)
 & \leq  \mu_{\epsilon}\\
  \label{eq:138}
\trace
\left(
\tilde{\cm{C}}\cm{P}^{-1}\tilde{\cm{C}}^T
\right)
& \leq \mu_{\eta}
.
\end{align}

Note that the upper bound of our objective function 
$c \tilde{\gamma}_{\epsilon}+\tilde{\gamma}_{\eta}$ is given by 
$c \sqrt{\mu_{\epsilon}}+\sqrt{\mu_{\eta}}$, which is not convex 
in $\mu_{\epsilon}$ and $\mu_{\eta}$. In stead of directly solving the problem, 
let us consider the following problem:
\begin{equation}
  \label{eq:136}
\min_{R[\mathrm z] \in RH_{\infty}, x_k \in \cm{E}(\cm{P}),\mu_{\epsilon}} \mu_{\epsilon}
\end{equation}
subject to $R[\infty]=1$, \eqref{eq:137} and \eqref{eq:138}.  

The condition $x_k \in \cm{E}(\cm{P})$ is described by \eqref{eq:130}, 
which is a bilinear matrix inequality (BMI) of the variables.  
On the other hand, by using the Schur complement,  
\eqref{eq:137} and \eqref{eq:138} 
can be expressed as linear matrix inequalities (LMIs): 
\begin{align}
  \label{eq:139}
        \left[
    \begin{array}{cc}
      \cm{P} &\cm{C}^T \\
      \cm{C} & \mu_{\epsilon}
    \end{array}
  \right]
&  \succeq \bm{0}
\\
  \label{eq:140}
        \left[
    \begin{array}{cc}
      \cm{P} &\tilde{\cm{C}}^T \\
      \tilde{\cm{C}} & \mu_{\eta}
    \end{array}
  \right]
&  \succeq \bm{0} 
.
\end{align}

Since the BMI is not convex, we cannot yet solve \eqref{eq:136}. 
Fortunately, we can convert the BMI into an LMI 
and then, since the LMI is convex, we can solve 
the minimization problem \eqref{eq:136} numerically 
as detailed in Appendix. 

By solving the convex optimization problem \eqref{eq:136}  
for different values for $\mu_{\eta}$, we obtain 
the optimal IIR feedback filters for different 
constraints on the $l_{\infty}$ norms of the feedback back signals. 
With the designed feedback filters $R[\mathrm z]-1$, 
we can evaluate the pair 
$(||R[\mathrm z]-1||, ||H[\mathrm z]R[\mathrm z]||)$   
and find the optimal $R[\mathrm z]$ that achieves the minimum of 
the left hand side of \eqref{eq:23}, which gives 
the minimum number for $b$. 

\subsection{Minimum $l_{\infty}$ norm for a fixed number of bits}
\label{section:minimum}
We have investigated the number of bits required for quantization 
to attain a prescribed requirement on the control performance
and the overloading-free
 quantization at the same time. 
Now, we would like to consider another problem to find
the achievable minimum $l_{\infty}$ norm of 
the error in the signal of interest 
with an overloading-free quantizer for a given number of bits. 
 
Suppose that the number of bits assigned to the static quantizer is 
given and fixed. 
We would like to design the overloading-free feedback quantizer 
that minimizes the $l_{\infty}$ norm of the error $\epsilon$. 
From \eqref{eq:19}, we obtain 
\begin{equation}
  \label{eq:41}
  L_y\leq  
  \left(
    2^{b-1}-\frac{1}{2}||R[\mathrm z]-1||
  \right)  d. 
\end{equation}
Since $d$ must be positive, we have to meet $||R[\mathrm z]-1||< 2^b$. 
It follows from $||\epsilon||_{\infty} \leq||H[\mathrm z]R[\mathrm z]||d/2$ 
that $||\epsilon||_{\infty}$ is bounded with $d=L_y/\left(
    2^{b-1}-||R[\mathrm z]-1||/2  \right)$ as 
  \begin{equation}
    \label{eq:42}
    ||\epsilon||_{\infty} \leq 
\frac{L_y ||H[\mathrm z]R[\mathrm z]||}{2^{b}-||R[\mathrm z]-1||}. 
  \end{equation}

It is obvious that for a fixed value of $||R[\mathrm z]-1||$, 
the upper bound for $||\epsilon||_{\infty}$ is minimized by the filter
$R[\mathrm z]$ that minimizes $||H[\mathrm z]R[\mathrm z]||$. 
Then, for a fixed upper bound $\tilde{\gamma}_{\eta}$ 
of $||R[\mathrm z]-1||$, the optimal filter can be found by 
solving the following optimization problem: 
\begin{equation}
  \label{eq:240}
  \min_{R[\mathrm z] \in RH_{\infty},\tilde{\gamma}_{\epsilon}}  \tilde{\gamma}_{\epsilon}
\end{equation}
subject to $R[\infty]=1$, \eqref{eq:25}, and \eqref{eq:26}. 

For a fixed upper bound 
for $||R[\mathrm z]-1||$, we have the value for 
$||H[\mathrm z]R[\mathrm z]||$ by solving the optimization problem. 
Then, by solving the problem for different values for $||R[\mathrm z]-1||$, 
the relationship 
between $||R[\mathrm z]-1||$ and $||H[\mathrm z]R[\mathrm z]||$ 
can be obtained. Finally, with the values for the pair 
$(||R[\mathrm z]-1||, ||H[\mathrm z]R[\mathrm z]||)$,  
we can obtain the minimum of the right hand side of \eqref{eq:42}. 

As in Sections \ref{section:FIR} and \ref{section:IIR},  
the problem can be formulated as an LP for FIR $R[\mathrm z]$, whereas 
the upper bounds of the norms have to be evaluated 
for IIR $R[\mathrm z]$. 
The details are omitted, since the optimization problems can be 
similarly solved as described in \ref{section:FIR} and \ref{section:IIR}.   

\section{Numerical examples} 
\begin{figure}
\begin{center}
 \centerline{\includegraphics[width=50mm]{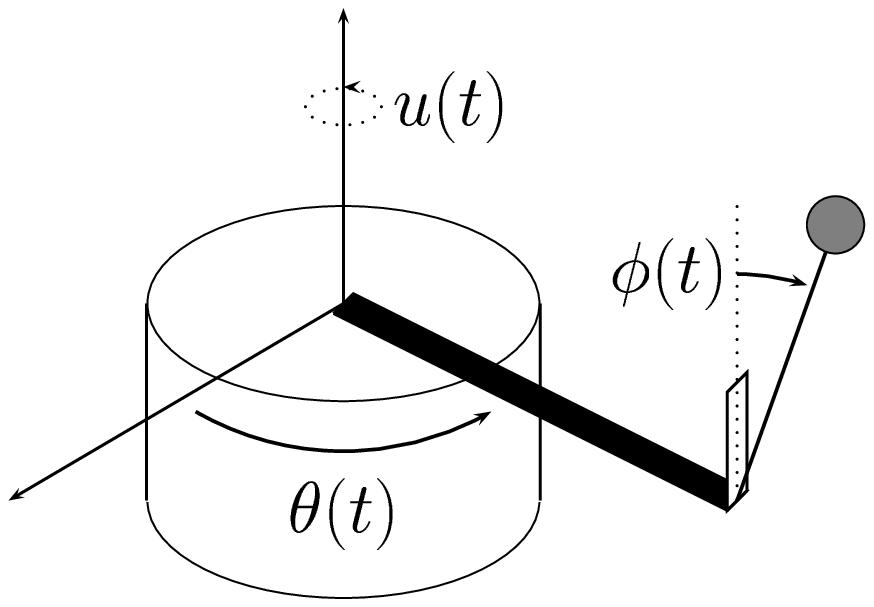}}
  \end{center}
\caption{Rotary inverted pendulum.}
\label{fig:pendulum}
\end{figure}

We here consider the rotary inverted pendulum (see e.g. \cite{1299533}) 
depicted in Fig. \ref{fig:pendulum} for our design example.

The pendulum connected at the end of the rotary arm is controlled 
by rotating the main body in the horizontal plane.  
The yaw angle of the arm is $\theta(t)$. 
The pendulum freely swings about a pitch angle $\phi(t)$  
in the vertical plane to the arm. 
The torque $u(t)$ is applied to actuate the pendulum. 
If $\phi(t)=0$, then the pendulum is balanced in the inverted position. 

We define the state of the rotary inverted pendulum as 
\begin{equation}
  \label{eq:43}
  x^T(t)=[ \phi(t), \theta(t), \dot{\phi}(t),\dot{\theta}(t)]
. 
\end{equation}
We periodically change the yaw angle, 
while keeping the stability of the rotary inverted pendulum.  
The target value of the yaw angle $\bar{\theta}(t)$ is 
\begin{equation}
  \label{eq:44}
\bar{\theta}(t)=
\left\{ \begin{array}{ll}
\frac{\pi}{2} & (10k \leq t < 5+10k) \\
0 & (5+10k \leq t < 10(k+1))
\end{array} \right. 
\end{equation}
for $k=0,1,\ldots$. 
The initial values of the states are assumed to be zero.

We linearize the continuous-time dynamical system of the pendulum and 
discretize this with the sampling period $T_s=0.01$.
Let $A,B,C,D$ be the state-space matrices of the linearized and discretized system.
Since the continuous-time system is strictly proper, we have $D=0$. 
The state-space matrices $A$ and $B$ of the discrete-time linearized system are given by 
\begin{eqnarray}
  \label{eq:45}
  A   &=&
  \left[
    \begin{array}{cccc}
    1.0056 &        0  &  0.0100  &  0.0001\\
   -0.0003 &   1.0000  & -0.0000  &  0.0100\\
    1.1134 &        0  &  1.0056  &  0.0149\\
   -0.0653 &        0  & -0.0003  &  0.9926
    \end{array}
  \right]
\\
B   &=&
  \left[
    \begin{array}{c}
   -0.0004\\
    0.0002\\
   -0.0864\\
    0.0431
    \end{array}
  \right]
.
\end{eqnarray}

Assuming that all of the state variables be available at the controller
(i.e. $C$ is the identity matrix), 
we adopt the state feedback control and determined its gain $K$
by the linear quadratic regulator (LQR) technique to  minimize 
\begin{equation}
  \label{eq:46}
\sum_{k=0}^{\infty}
\left(
x_k^T Q_{lqr} x_k + r |u_k|^2  
\right)
\end{equation}
where the weights are 
\begin{equation}
  \label{eq:47}
Q_{lqr}= \diag [10, 2, 0.5, 0], \quad r=0.05.  
\end{equation}

Our signals of interest is the  discretized $\phi(t)$, 
which is expressed as $\phi_k=C_1 x_k$ 
with 
\begin{align}
  \label{eq:48}
  C_1  &=
  \left[
    \begin{array}{cccc}
     1  &   0  &   0 &    0
    \end{array}
  \right]
.
\end{align}
The transfer function from 
the $l$th entry of the quantization error $\eta$ to $\phi $ and $\theta$ 
is found to be $C_1(\mathrm z I-A-BK)^{-1}BK_l$ 
with $K=[K_1,K_2,K_3,K_4]=[57.2598, 6.0910, 6.2562,3.4953]$. 

Now let us design FIR and IIR filters of order 4 for the error feedback.  
We consider the quantization of $\phi$, 
the first entry of the state variables,  
to mitigate the effect of the quantization on $\phi$. 
The transfer function is given by 
\begin{align}
  \label{eq:49}
H[\mathrm z]&=
C_1(\mathrm z I-A-BK)^{-1}BK_1\\
\nonumber
& =
\frac{0.02475 \mathrm z^3 - 0.02482 \mathrm z^2 - 0.02463 \mathrm z + 0.02469}
{\mathrm z^4 - 3.59 \mathrm z^3 + 4.808 \mathrm z^2 - 2.844 \mathrm z + 0.626}
\end{align}
whose zeros are $-0.9975$, $1$, and $1$. 

The constraint on the maximum absolute value of 
$\epsilon$ is set to be $\gamma_{\epsilon}=0.05$ and 
$L_y$ in \eqref{eq:13} is set to be $\pi/2$. 

With the designed optimal FIR feedback filter, 
the value of the objective function 
$c||H[\mathrm z]R[\mathrm z]||+||R[\mathrm z]-1||$ is 5.2729.  
On the other hand, the value with the designed IIR feedback filter is 5.8831. 
The FIR filter exhibits a  better performance than the IIR filter. 
This is due to the fact that the exact value of the $l_\infty$ norm is evaluated 
for the design of the FIR filter,  
whereas only an upper bound can be used 
for the design of the IIR filter.   
In both cases, the required number of bits is 3
to satisfy the constraint \eqref{eq:23},
while the conventional static uniform quantizer requires 
$6$ bits  to meet the constraint on $\epsilon$, 
since $(L_y/\gamma_{\epsilon})||H[\mathrm z]||=47.788< 2^6$. 

Simulations are conducted with the designed optimal FIR error feedback filter. 
To clarify the difference, we only quantize the signal $\phi$ of interest.   

\begin{figure}[t!]
\begin{center}
 \centerline{\includegraphics[width=80mm]{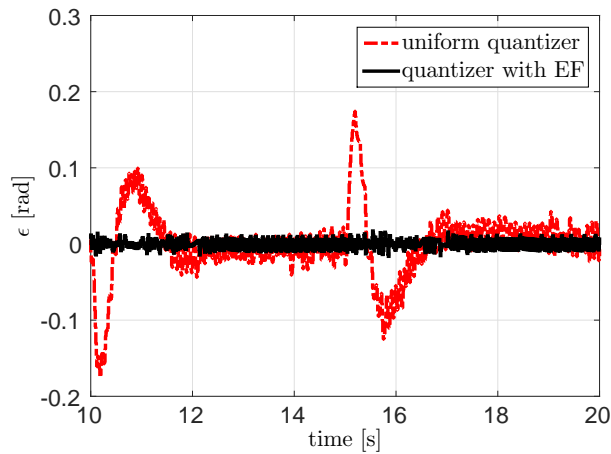}}
 \caption{Error of pitch angles of the pendulum controlled 
with the $3$-bit quantizers having the optimized FIR 
error feedback filter (solid line) 
and the conventional uniform quantizers (dash-dotted line). 
}
 \label{fig:output}
  \end{center}
\end{figure}

\begin{figure}[t!]
\begin{center}
 \centerline{\includegraphics[width=80mm]{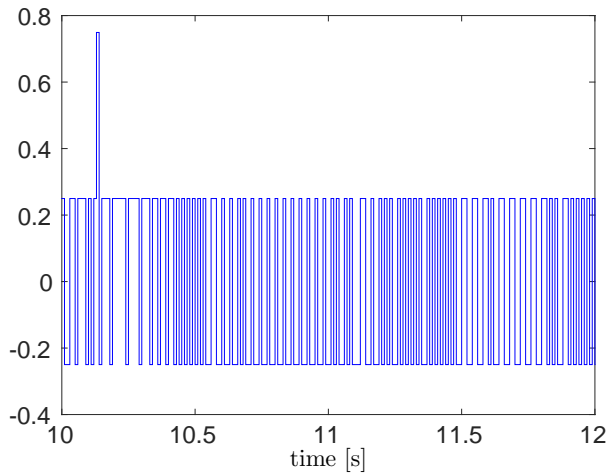}}
 \caption{Output of the designed quantizer for pitch angle.}
 \label{fig:quan_sig}
  \end{center}
\end{figure}

Fig. \ref{fig:output} compares the error signal $\epsilon$ 
of the pendulum controlled with the $3$-bit quantizers having  
the optimized FIR error feedback filter (solid line) and  
the conventional uniform quantizer (dash-dotted line) for $10\leq t<20$. 
The maximum absolute value of the error for our designed quantizer 
is less than 0.05, while the maximum absolute value of the error 
for the conventional uniform quantizers is about 0.18. 
Our designed quantizer satisfies the requirement on the error  
clearly outperforms the conventional uniform quantizer. 

The output of our designed quantizer for $10\leq t<12$ 
is shown in Fig. \ref{fig:quan_sig}. 
Only three values are taken, which implies that only 2 bits are 
required in practice, although our analysis suggests 3 bits. 
This is because we adopt the worst-case error for our performance measure.  
Indeed, it is well-known that the condition on the maximum of 
the absolute value of errors leads to conservative results. 

\begin{figure}[t!]
\begin{center}
 \centerline{\includegraphics[width=80mm]{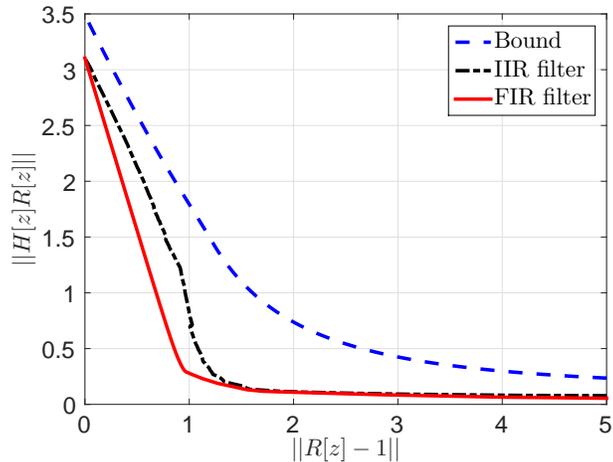}}
 \caption{Relation between 
$||R[\mathrm z]-1||$ and 
$||H[\mathrm z]R[\mathrm z]||$ for pitch angle $\phi$.}
 \label{fig:FIR_IIR_pendulum}
  \end{center}
\end{figure}

Next, for a fixed number of bits for the quantizer, 
we evaluate the $l_{\infty}$ norm of the error in the signal of interest 
with the designed overloading-free quantizer. 
We solve the optimization problems discussed in Section \ref{section:minimum}.  

Fig. \ref{fig:FIR_IIR_pendulum} depicts 
$||H[\mathrm z]R[\mathrm z]||$ as a function of $||R[\mathrm z]-1||$. 
In the design of IIR filters, we minimize the upper bound and 
then the designed filter is not assured to be optimal. 
Here, $(\sqrt{\mu_{\eta}}, \sqrt{\mu_{\epsilon}})$ serves as 
an upper bound for 
$(||R[\mathrm z]-1||,||H[\mathrm z]R[\mathrm z]||)$ of IIR filters.  
On the other hand, in the design of FIR filters, 
we minimize the objective function directly and 
the designed filter is optimal among FIR filters. 
This may be a reason why the designed FIR filters achieve 
smaller error norm than the designed IIR filers. 

As $||R[\mathrm z]-1||$ increases from zero, 
$||H[\mathrm z]R[\mathrm z]||$ decreases rapidly at first and 
then floors. 
It should be remarked that $||R[\mathrm z]-1||=0$ implies 
that there is not any error feedback filter, that is, 
the quantizer is just a static uniform quantizer. 


\begin{figure}[t!]
\begin{center}
 \centerline{\includegraphics[width=80mm]{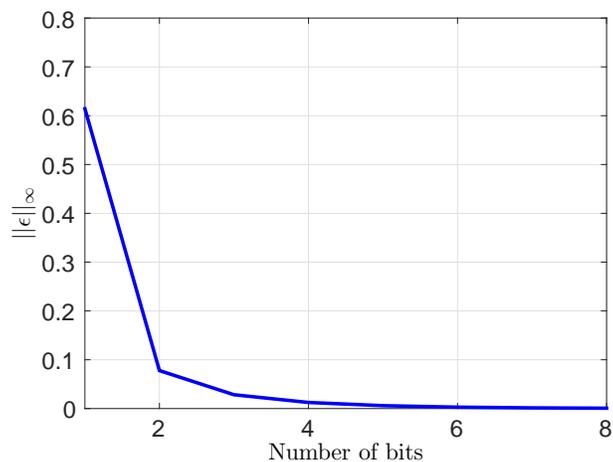}}
 \caption{$||\epsilon||_{\infty}$ for different numbers of bits.}
 \label{fig:bit_vs_bound}
  \end{center}
\end{figure}

From the values of $(||R[\mathrm z]-1||, ||H[\mathrm z]R[\mathrm z]||)$ 
in Fig. \ref{fig:FIR_IIR_pendulum}, 
we compute the norm $||\epsilon||_{\infty}$ 
with \eqref{eq:17} for different numbers of bits from 1 to 8, 
which is plotted in Fig. \ref{fig:bit_vs_bound}. 
This figure clarifies the relationship between 
the error norm and the number of bits assigned to the quantizer. 

The norm of our quantizer decays exponentially with a rate 
faster than $1/2$. 
We may conclude that our quantizer is more efficient 
in the number of bits 
than the conventional quantizer without the error feedback filter, 
since its decay rate of the error norm with respect to 
the number of bits is given by $1/2$.

\section{Conclusion}
We have studied a feedback quantizer composed of 
a static quantizer and an error feedback filter. 
It has been 
shown that quantizers can be designed independently of the control law. 
Then, we have investigated 
the necessary number of bits required for quantization 
to attain the requirement on the system output, while keeping 
no-overloading in the quantizer. 
The number of bits assigned to the quantizer can be obtained 
by designing the error feedback filter that minimizes 
a constraint for no-overloading. 
The design of FIR filters has been formulated as linear programming 
by directly evaluating the $l_{\infty}$ norm, 
whereas the design of IIR filters has been as a convex optimization problem 
by using upper bounds on the $l_{\infty}$ norm.   
In our design example, if one assigns the same order 
for filters, 
the optimized FIR filter exhibits a better performance than 
the designed IIR filter.  
The efficiency of the designed quantizer has been demonstrated 
by simulation.


\begin{thebibliography}{10}
\providecommand{\url}[1]{#1}
\csname url@samestyle\endcsname
\providecommand{\newblock}{\relax}
\providecommand{\bibinfo}[2]{#2}
\providecommand{\BIBentrySTDinterwordspacing}{\spaceskip=0pt\relax}
\providecommand{\BIBentryALTinterwordstretchfactor}{4}
\providecommand{\BIBentryALTinterwordspacing}{\spaceskip=\fontdimen2\font plus
\BIBentryALTinterwordstretchfactor\fontdimen3\font minus
  \fontdimen4\font\relax}
\providecommand{\BIBforeignlanguage}[2]{{%
\expandafter\ifx\csname l@#1\endcsname\relax
\typeout{** WARNING: IEEEtran.bst: No hyphenation pattern has been}%
\typeout{** loaded for the language `#1'. Using the pattern for}%
\typeout{** the default language instead.}%
\else
\language=\csname l@#1\endcsname
\fi
#2}}
\providecommand{\BIBdecl}{\relax}
\BIBdecl

\bibitem{1299533}
N.~Ploplys, P.~Kawka, and A.~Alleyne, ``Closed-loop control over wireless
  networks,'' \emph{IEEE Control Systems}, vol.~24, no.~3, pp. 58--71, Jun
  2004.

\bibitem{wong1999systems}
W.~S. Wong and R.~W. Brockett, ``Systems with finite communication bandwidth
  constraints. {II}. stabilization with limited information feedback,''
  \emph{IEEE Transactions on Automatic Control}, vol.~44, no.~5, pp.
  1049--1053, 1999.

\bibitem{Tatikonda2004}
S.~Tatikonda and S.~Mitter, ``{Control Under Communication Constraints},''
  \emph{IEEE Transactions on Automatic Control}, vol.~49, no.~7, pp.
  1056--1068, Jul. 2004. 

\bibitem{Nair2003}
G.~N. Nair and R.~J. Evans, ``{Exponential stabilisability of
  finite-dimensional linear systems with limited data rates},''
  \emph{Automatica}, vol.~39, no.~4, pp. 585--593, Apr. 2003. [Online].

\bibitem{1162925}
Tran-Thong and B.~Liu, ``Error spectrum shaping in narrow-band recursive
  filters,'' \emph{IEEE Transactions on Acoustics, Speech and Signal
  Processing}, vol.~25, no.~2, pp. 200--203, Apr 1977.

\bibitem{1163989}
W.~Higgins and D.~Munson, ``Noise reduction strategies for digital filters:
  Error spectrum shaping versus the optimal linear state-space formulation,''
  \emph{IEEE Transactions on Acoustics, Speech and Signal Processing}, vol.~30,
  no.~6, pp. 963--973, Dec 1982.

\bibitem{134473}
T.~Laakso and I.~Hartimo, ``Noise reduction in recursive digital filters using
  high-order error feedback,'' \emph{IEEE Transactions on Signal Processing},
  vol.~40, no.~5, pp. 1096--1107, May 1992.

\bibitem{Delta-Sigma}
R.~Schreier and G.~C. Temes, \emph{Understanding Delta-Sigma Data
  Converters}.\hskip 1em plus 0.5em minus 0.4em\relax Wiley-IEEE Press, 2004.

\bibitem{Azuma2008b}

S.~Azuma and T.~Sugie, ``{Optimal dynamic quantizers for discrete-valued input
  control},'' \emph{Automatica}, vol.~44, no.~2, pp. 396--406, Feb. 2008.

\bibitem{Azuma2008}
------, ``Synthesis of optimal dynamic quantizers for discrete-valued input
  control,'' \emph{IEEE Transactions on Automatic Control}, vol.~53, no.~9, pp.
  2064--2075, Oct. 2008. 

\bibitem{Sawada2010c}
K.~Sawada and S.~Shin, ``Dynamic quantizer synthesis based on invariant set
  analysis for {SISO} systems with discrete-valued input,'' in \emph{the 19th
  International Symposium on Mathematical Theory of Networks and Systems},
  2010, pp. 1385--1390.

\bibitem{7769239}
S.~Ohno and M.~R. Tariq, ``Optimization of noise shaping filter for quantizer
  with error feedback,'' \emph{IEEE Transactions on Circuits and Systems I:
  Regular Papers}, vol.~PP, no.~99, pp. 1--13, 2016.

\bibitem{6156470}
M.~Nagahara and Y.~Yamamoto, ``Frequency domain min-max optimization of
  noise-shaping delta-sigma modulators,'' \emph{IEEE Transactions on Signal
  Processing}, vol.~60, no.~6, pp. 2828--2839, June 2012.

\bibitem{li2014design}
X.~Li, C.~B. Yu, and H.~Gao, ``Design of delta--sigma modulators via
  generalized {K}alman--{Y}akubovich--{P}opov lemma,'' \emph{Automatica},
  vol.~50, no.~10, pp. 2700--2708, 2014.

\bibitem{6461108}
S.~Callegari and F.~Bizzarri, ``Output filter aware optimization of the noise
  shaping properties of {$\Delta$} {$\Sigma$} modulators via semi-definite
  programming,'' \emph{IEEE Transactions on Circuits and Systems I: Regular
  Papers}, vol.~60, no.~9, pp. 2352--2365, Sept 2013.

\bibitem{6617704}
------, ``Noise weighting in the design of {$\Delta$} {$\Sigma$} modulators
  (with a psychoacoustic coder as an example),'' \emph{IEEE Transactions on
  Circuits and Systems II: Express Briefs}, vol.~60, no.~11, pp. 756--760, Nov
  2013.

\bibitem{4522532}
M.~Derpich, E.~Silva, D.~Quevedo, and G.~Goodwin, ``On optimal perfect
  reconstruction feedback quantizers,'' \emph{IEEE Transactions on Signal
  Processing}, vol.~56, no.~8, pp. 3871--3890, Aug 2008.

\bibitem{ohno2017eusipco}
S.~Ohno, M.~R. Tariq, and M.~Nagahara, ``Min-max {IIR} filter design for
  feedback quantizers,'' submitted to EUSIPCO 2017, Feb. 2017.

\bibitem{832826}
V.~Chellaboina, M.~Haddad, D.~Bernstein, and D.~Wilson, ``Induced convolution
  operator norms for discrete-time linear systems,'' in \emph{Proceedings of
  the 38th IEEE Conference on Decision and Control, 1999.}, vol.~1, 1999, pp.
  487--492 vol.1.

\bibitem{Shingin}
H.~Shingin and Y.~Ohta, ``Optimal invariant sets for discrete-time systems:
  Approximation of reachable sets for bounded inputs,'' in \emph{10th
  IFAC/IFORS/IMACS/IFIP Symposium on Large Scale Systems: Theory and
  Applications (LSS)}, 2004, pp. 401--406.

\bibitem{599969}
C.~Scherer, P.~Gahinet, and M.~Chilali, ``Multiobjective output-feedback
  control via {LMI} optimization,'' \emph{IEEE Transactions on Automatic
  Control}, vol.~42, no.~7, pp. 896--911, Jul 1997.

\bibitem{Masubuchi98}
I.~Masubuchi, A.~Ohara, and N.~Suda, ``{LMI}-based controller synthesis: {A}
  unified formulation and solution,'' \emph{International Journal of Robust and
  Nonlinear Control}, vol.~8, no.~8, p. 669–686, July 1998.

\bibitem{cvx}
M.~Grant and S.~Boyd, ``{CVX}: Matlab software for disciplined convex
  programming, version 2.0 beta,'' \url{http://cvxr.com/cvx}, Sep. 2012.

\end{thebibliography}

\appendix
\section{Numerical evaluation based on LMIs}
\label{sec:appen1}

Let us convert the non-convex BMI 
\eqref{eq:130}  to an LMI by using the change of variables 
proposed independently in \cite{599969} and \cite{Masubuchi98}. 

Let the order of $R[\mathrm z]$ be equal to 
the order $n$ of the system $H[\mathrm z]$. 
The set of $n\times n$ positive definite matrices 
is denoted as $PD(n)$. We  define the following matrices 
$\{P_f, S_f, W_f, W_g, L\}$, where 
$P_f \in PD(n)$, $S_f \in PD(n)$, $W_f \in \mathbb{R}^{1 \times n}$, 
$W_g \in \mathbb{R}^{n \times 1}$, $L \in \mathbb{R}^{n \times n }$, 
with $P_f$ and $P_g$. 
Let us also define matrices from $\{P_f, S_f, W_f, W_g, L\}$ 
as 
\begin{eqnarray}
  \label{eq:51}
  \cm{P}^{-1}&=&
      \left[
    \begin{array}{cc}
       P_f & S_f \\ 
       S_f & S_f 
    \end{array}
  \right]
\\
U&=&
      \left[
    \begin{array}{cc}
       P_f& I_n \\
       S_f & \bm{0}
    \end{array}
  \right]
\\
  \label{eq:52}
  P_g&=&(P_f-S_f)^{-1}
\end{eqnarray}
and the matrices $\{M_{\cm{A}}, M_{\cm{B}}, M_{\cm{C}},M_{\cm{P}}\}$ as
\begin{align}
\label{eq:53}
M_{\cm{A}}&=
\left[
  \begin{array}{cc}
    A_hP_f +B_h W_{f} & A_h \\
    L & P_g A_h  \\
  \end{array}
\right]
\\ \label{eq:54}
M_{\cm{B}}&=
\left[
  \begin{array}{c}
    B_h\\
    W_g
  \end{array}
\right]
\\ 
\label{eq:55}
M_{\cm{C}}&=
\left[
  \begin{array}{cc}
    C_hP_f + D_hW_f& C_h
  \end{array}
\right]
\\ \label{eq:56}
  M_{\cm{P}}&=
  \left[
    \begin{array}{cc}
      P_f & I_n \\ 
      I_n & P_g \\
    \end{array}
  \right]
\end{align}
Direct computations show that 
if the matrices $\{A_r, B_r, C_r\}$ are 
\begin{align}
  \label{eq:57}
  A_r&=[B_hW_{f}-P_g^{-1}(L-P_g A_h P_f)]S_f^{-1} 
\\
  \label{eq:58}
B_r&=B_h-P_g^{-1}W_{g} 
\\
  \label{eq:59}
C_r&=W_{f} S_f^{-1} 
\end{align}
then $\{\cm{A}, \cm{B}, \cm{C}\}$ satisfy 
\begin{eqnarray}
  \label{eq:60}
M_{\cm{A}} &=& U^T \cm{P}\cm{A}U\\
M_{\cm{B}} &=&U^T\cm{P} \cm{B}\\
M_{\cm{C}} &=&\cm{C}U\\
M_{\cm{P}}&=& U^T \cm{P} U
.
\end{eqnarray}
Theorem 1 \cite{Masubuchi98} proves that the BMI 
for the original variables $\{\cm{A}, \cm{B}, \cm{C}, \cm{P}\}$
is equivalent to the LMI for the new variables $\{M_{\cm{A}},M_{\cm{B}},M_{\cm{C}},M_{\cm{P}}\}$ by replacing  $\{\cm{P}\cm{A}, \cm{P}\cm{B},\cm{C}, \cm{P}\}$ 
with $\{M_{\cm{A}},M_{\cm{B}},M_{\cm{C}},M_{\cm{P}}\}$. 
Thus, \eqref{eq:130} and \eqref{eq:139} are converted into 
\begin{align}
  \label{eq:61}
      \left[
    \begin{array}{ccc}
      (1-\alpha)M_{\cm{P}} &\bm{0}&M_{\cm{A}}^T\\
      \bm{0} & \alpha &M_{\cm{B}}^T\\
      M_{\cm{A}} &M_{\cm{B}}& M_{\cm{P}}
    \end{array}
  \right]
  &\succeq \bm{0}
\\
  \label{eq:62}
        \left[
    \begin{array}{cc}
      M_{\cm{P}} &M_{\cm{C}}^T \\
      M_{\cm{C}} & \mu_{\epsilon}
    \end{array}
  \right]
&  \succeq \bm{0}.
\end{align}

On the other hand, we have 
\begin{equation}
  \label{eq:63}
  \tilde{\cm{C}}U=
  \left[
    \begin{array}{cc}
      C_rS_f  & \bm{0} 
    \end{array}
  \right]
=
  \left[
    \begin{array}{cc}
      W_f  & \bm{0} 
    \end{array}
  \right]
:=M_{\tilde{\cm{C}}}
.
\end{equation}
Premultiplying \eqref{eq:140} by $\diag(U^T,I)$  and 
postmultiplying \eqref{eq:140} by $\diag(U,I)$ results in 
\begin{equation}
  \label{eq:64}
        \left[
    \begin{array}{cc}
      M_{\cm{P}} &M_{\tilde{\cm{C}}}^T \\
      M_{\tilde{\cm{C}}}  &\mu_{\eta}
    \end{array}
  \right]
  \succeq \bm{0}.
\end{equation}
Therefore the minimization problem 
\begin{equation}
  \label{eq:65}
\min_{P_{f}, P_{g},W_f, W_g, L, \mu_{\epsilon}}  \mu_{\epsilon}
\end{equation}
subject to \eqref{eq:61}, \eqref{eq:62}, and \eqref{eq:64}, gives 
the minimum of the  minimization problem \eqref{eq:136} 
for a given $\alpha$. 

For a fixed $\alpha$, 
the minimization problem is a semidefinite program, 
which can be numerically solved by existing 
optimization packages, e.g., CVX \cite{cvx}, 
a package for specifying and solving convex programs.  
then, the minimum is given by a line search for $\alpha \in (0,\rho^2(A_h))$.

\end{document}